\documentclass[a4paper,12pt]{article}
\usepackage{amsmath,amsfonts,bm,amssymb,verbatim} 
\usepackage{inputenc}
\usepackage{xcolor,colortbl}
\usepackage[english]{babel}
\usepackage{graphicx}
\usepackage{float}
\usepackage{verbatim}
\usepackage{subfigure}
\usepackage{lscape}
\usepackage{enumerate}
\usepackage{setspace}
\usepackage{multirow}
\usepackage{rotating}
\RequirePackage[colorlinks]{hyperref}

\newcommand{\bes}{\begin{equation*}}
\newcommand{\ees}{\end{equation*}}
\newcommand{\be}{\begin{equation}}
\newcommand{\ee}{\end{equation}}

\newcommand{\ue}{\mathrm{e}}

\begin{document}

\title{Collective synchronization and high frequency systemic instabilities in financial markets}

\author{%
Lucio Maria Calcagnile$^{\textrm{a,b,}}$\footnote{Corresponding author. \textit{E-mail address}: l.calcagnile@list-group.com}~,
Giacomo Bormetti$^{\textrm{c,b}}$,
Michele Treccani$^{\textrm{a,b,}}$\thanks{Present address: Mediobanca S.p.A, Piazzetta E.~Cuccia 1, 20121 Milano, Italy}~,\\
Stefano Marmi$^{\textrm{c,d}}$, and Fabrizio Lillo$^{\textrm{c,e}}$
}
\date{\today{}}

\maketitle

\small
\begin{center}
  $^\textrm{a}$~\emph{LIST S.p.A., via Pietrasantina 123, 56122 Pisa, Italy}\\
  $^\textrm{b}$~\emph{QUANTLab\hspace{2pt}\footnote{www.quantlab.it}, via Pietrasantina 123, Pisa, 56122, Italy}\\
  $^\textrm{c}$~\emph{Scuola Normale Superiore, Piazza dei Cavalieri 7, Pisa, 56126, Italy}\\
  $^\textrm{d}$~\emph{CNRS UMI 3483 – Laboratorio Fibonacci, Piazza dei Cavalieri 7, Pisa, 56126, Italy}\\
  $^\textrm{e}$~\emph{Santa Fe Institute, 1399 Hyde Park Road, Santa Fe, NM 87501, USA}\\
\end{center}
\normalsize

\begin{abstract}
  \noindent Recent years have seen an unprecedented rise of the role that technology plays in all aspects of human activities. Unavoidably, technology has heavily entered the Capital Markets trading space, to the extent that all major exchanges are now trading exclusively using electronic platforms. The ultra fast speed of information processing, order placement, and cancelling generates new dynamics which is still not completely deciphered. Analyzing a large dataset of stocks traded on the US markets, our study evidences that since 2001 the level of synchronization of large price movements across assets has significantly increased. Even though the total number of over-threshold events has diminished in recent years, when an event occurs, the average number of assets swinging together has increased. Quite unexpectedly, only a minor fraction of these events -- regularly less than 40\% along all years -- can be connected with the release of pre-announced macroeconomic news. We also document that the larger is the level of sistemicity of an event, the larger is the probability -- and degree of sistemicity -- that a new event will occur in the near future. This opens the way to the intriguing idea that systemic events emerge as an effect of a purely endogenous mechanism. Consistently, we present a high-dimensional, yet parsimonious, model based on a class of self- and cross-exciting processes, termed Hawkes processes, which reconciles the modeling effort with the empirical evidence.
\end{abstract}

\section{Introduction}

Quoting from Michael Lewis' Flash Boys ``The world clings to its old mental picture of the stock market because it's comforting''~\cite{lewis2014flash}. But trading activity has profoundly changed from the old phone conversation or click and trade on a screen to software programming. Market statistics confirm that automated algorithms carry out a significant fraction of the trading activity on US and Europe electronic exchanges~\cite{gomber2011high,macintosh2013high}. As algos feed on financial and news data, the speed of information processing has dramatically increased and potentially allows large price movements to propagate very rapidly through different assets and exchanges~\cite{gerig2013high}. 

The synchronization effect had its most spectacular appearance during the May 6th, 2010 Flash Crash. The crash started from a rapid price decline in the E-Mini S\&P 500 market and in a very short time the anomaly became systemic and the shock propagated towards ETFs, stock indices and their components, and derivatives~\cite{sec2010,kirilenko2014flash}. The price of the Dow Jones Industrial Average plunged by 9\% in less than 5 minutes but recovered the pre-shock level in the next 15 minutes of trading. The SEC reported that such a swing was sparked by an algorithm executing a sell order placed by a large mutual fund. Then high frequency traders, even though did not ignited the event, caused a ``hot potato'' effect amplifying the crash. 
In the aftermath of the crash, several studies have focused on events, evocatively named \textit{Mini Flash Crashes}, concerned with the emergence of large price movements of an asset in a very limited fraction of time
and attributing their origin to the interaction between several automatic algorithms~\cite{johnson2013abrupt} or to the unexpected product of regulation framework and market fragmentation~\cite{golub2012high}.

The Flash Crash, however, has also dramatically shown how strongly interconnected different markets and asset classes can become, especially during extreme events. In this paper, by taking a different, yet complementary approach to the above literature, we investigate how the frequency of collective instabilities at high frequency has changed in the last years. Specifically, we identify one-minute extreme events as over-threshold movements. In this respect, our approach shares some similarities with previous works employing non-parametric tests to identify extreme movements, see~\cite{bormetti2015modelling,andersen2007no,lee2008jumps, andersen2010continuous,dumitru2012identifying}. We perform our analysis on a yearly basis from 2001 to 2013 on a data sample of highly liquid US equities and we identify extreme events  affecting a sizable fraction of the investigated assets. Remarkably, very little research has been devoted to the investigation of this kind of systemic events. Few noticeable exceptions are~\cite{bollerslev2013jump}, who aim at the identification of common large movements between the market portfolio and individual stocks, and~\cite{gilder2014cojumps}, who investigate the tendency of large movements to arrive simultaneously. A very recent non-parametric test of the occurrence of simultaneous jumps across multiple assets is discussed in~\cite{caporin2014multi}. Our research provides the empirical evidence that, while the total number of extreme movements has decreased along years, the occurrence of systemic events has significantly increased. 

To identify the possible causes of such events we compare their time occurrences with a database of pre-scheduled macroeconomic announcements. Since macroeconomic news can be expected to have a market-level influence, they represent a natural candidate to explain market-wide events. For instance, literature has recognized the peculiar role played by Federal Open Market Committee (FOMC) meetings deciding the interest rate level~\cite{petersen2010quantitative,petersen2010market}. However, unexpectedly, only a minor fraction (less than 40\%) of events involving a large fraction of assets has been preceded by the release of a macro news. This evidence opens the route to the more intriguing hypothesis that a genuinely endogenous dynamics is taking place. To the best of our knowledge, the association between extreme equity price movements and the news arrival has been previously investigated in~\cite{lee2008jumps,lee2011jumps}, finding a positive association, but the results have been challenged in~\cite{bajgrowicz2013jumps}. Table 11~in~\cite{gilder2014cojumps} suggests the existence of a particularly strong relationship between FOMC announcements and the arrival of a systemic event (defined as an event when the market index jumps). However, none of the previous works performs an analysis of the association between news and extreme movements conditional on the level of systemicity of the event. 

Finally, we show that when an event affecting a significant fraction of assets occurs, the probability of a novel extreme event in the subsequent minutes increases. More interestingly, there is a clear evidence that the more systemic the conditioning event is, the larger the expected number of assets swinging synchronously in the immediate future will be. In order to reproduce such empirical evidences, we propose a model within the class of mutually exciting point processes, termed Hawkes processes~\cite{Hawkes:1971} which in recent years have experienced an increasing popularity in mathematical finance and econometrics~\cite{bowsher2007modelling,bauwens2009modelling,Muni-Toke:2011,Muni-Toke_Pomponio:2012,Filimonov_Sornette:2012,Bacry_etal:2013,hardiman2013critical,rambaldi2014modeling,Bacry_etal:2015}. We present a multidimensional, yet parsimonious, Hawkes process which captures with remarkable realism the cross-excitation affecting over-threshold events.

\section{Data} \label{sec:data}

{\bf Financial data.} We conduct our analysis on price time series of financial stocks belonging to the Russell 3000 Index, traded in the US equity markets (mostly NYSE and NASDAQ). We consider the thirteen years from 2001 to 2013 and for each year we select 140 highly liquid stocks. 
We use 1-minute closing price data during the regular US trading session, i.e.~from 9:30 a.m.~to 4:00 p.m and, as explained in the \emph{Support Information}, we remove the intraday pattern of volatility, which is a local measure of the diffusion rate of price. 

{\bf News data.}
We use macroeconomic news data provided by Econoday, Inc. \textit{www.econoday.com}. We consider the 42 most important news categories, which are classified into two large groups according to their capacity of influencing the financial markets: the Market Moving Indicator group and the Merit Extra Attention group. Since we are concerned with matching news with market extreme events, we consider only the 27 categories whose announcement times occur during the trading session. The number of total news announcements ranges from around 150 in the first years to around 260 in the last years, for a total of 2,888 news. See the \emph{Support Information} for more details.

\section{Methods} \label{sec:methods}
\subsection{Identification of extreme events}
In order to detect extreme variations of the stock prices $P_t$, we compare price returns (defined as $r_t = \ln {P_t}/P_{t-1}$) with an estimate of the historical spot volatility, which sets the scale of local price fluctuations. Specifically, we calculate a volatility time series $\sigma_t$ as an exponential-moving-average version of the bipower variation (see~\cite{bormetti2015modelling,Barndorff-Nielsen_Shephard:2004,Corsi_etal:2010}) of the return time series and we finally say that an extreme return occurs when
\begin{equation}\label{eq:jump}
	\frac{|r_t|}{\sigma_t} > \theta ,
\end{equation}
for a certain threshold $\theta$. In our main analyses we take $\theta = 4$, but we also investigate higher values of the threshold, namely $\theta = 6$, 8, 10, in some of our descriptive statistics.

\section{Results} \label{sec:results}
The main objective of this paper is the modeling of the dynamics of synchronous large price variations at high frequency. We say that a stock {\it jumps} in a given one minute interval if condition of Eq.~\ref{eq:jump} is observed for a given $\theta$. Here we are mostly interested in {\it cojumps}, i.e. the simultaneous (inside the minute) occurrence of jumps for a subset of $M$ stocks. The quantity $M$ is termed the {\it multiplicity} of the cojump, and it gives a measure of the systemic nature of the event. In the following we consider three questions: (i) how has the high frequency instability changed in the last fifteen years? (ii) what fraction of the systemic instabilities can be attributed to macroeconomic news? (iii) how can we model the short term dynamics of market instabilities?

\subsection{Historical dynamics of jumps and cojumps}
\begin{figure*}[t]
  \centering
  \includegraphics[width=\textwidth]{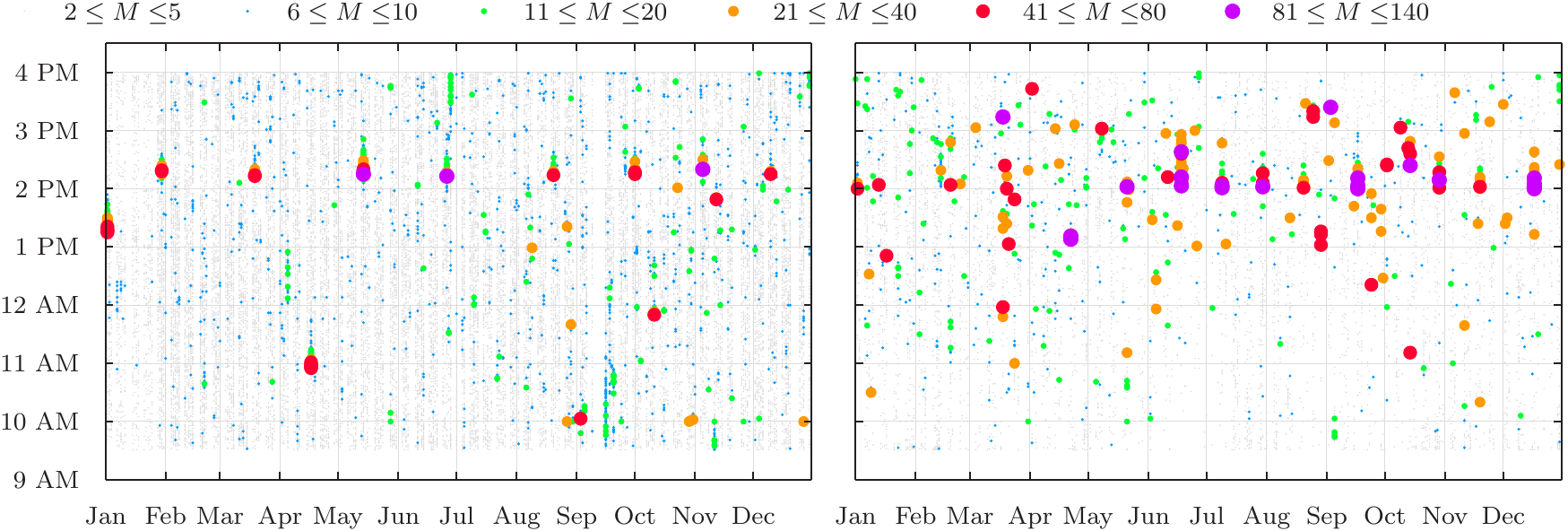}
  \caption{Time series of the cojumps detected for the dataset of 140 selected highly liquid stocks of the Russell 3000 Index during year 2001 (left panel) and 2013 (right panel). The size of the circles increases with the multiplicity of the cojump event.} \label{fig:cojumps}
\end{figure*}
A visual representation of how instability of financial markets has changed in the last years is shown in Fig.~\ref{fig:cojumps}, which compares the dynamics of $\theta=4$ cojumps in 2001 (left panel) and 2013 (right panel). The horizontal axis represents the trading day and the vertical axis indicates the hour of the day. The presence of a circle indicates the occurrence of a cojump and the color codifies the number of stocks simultaneously cojumping (i.e.~the multiplicity). In 2001 there were many cojumps with low multiplicity and the high multiplicity cojumps are concentrated mostly at specific hours of the day (10 a.m.~and 2:15 p.m.) corresponding to the release of important macro announcements, such as, for example, the FOMC announcements. On the contrary, in 2013 we observe less low multiplicity cojumps and many more high multiplicity cojumps, which are quite scattered during the day. This is an indication that modern financial markets have become more systemically unstable and that these instabilities are less related to macro news. In the following we show that this is the case with more quantitative analyses.

\begin{figure*}[t!]
  \centering
  \includegraphics[width=\textwidth]{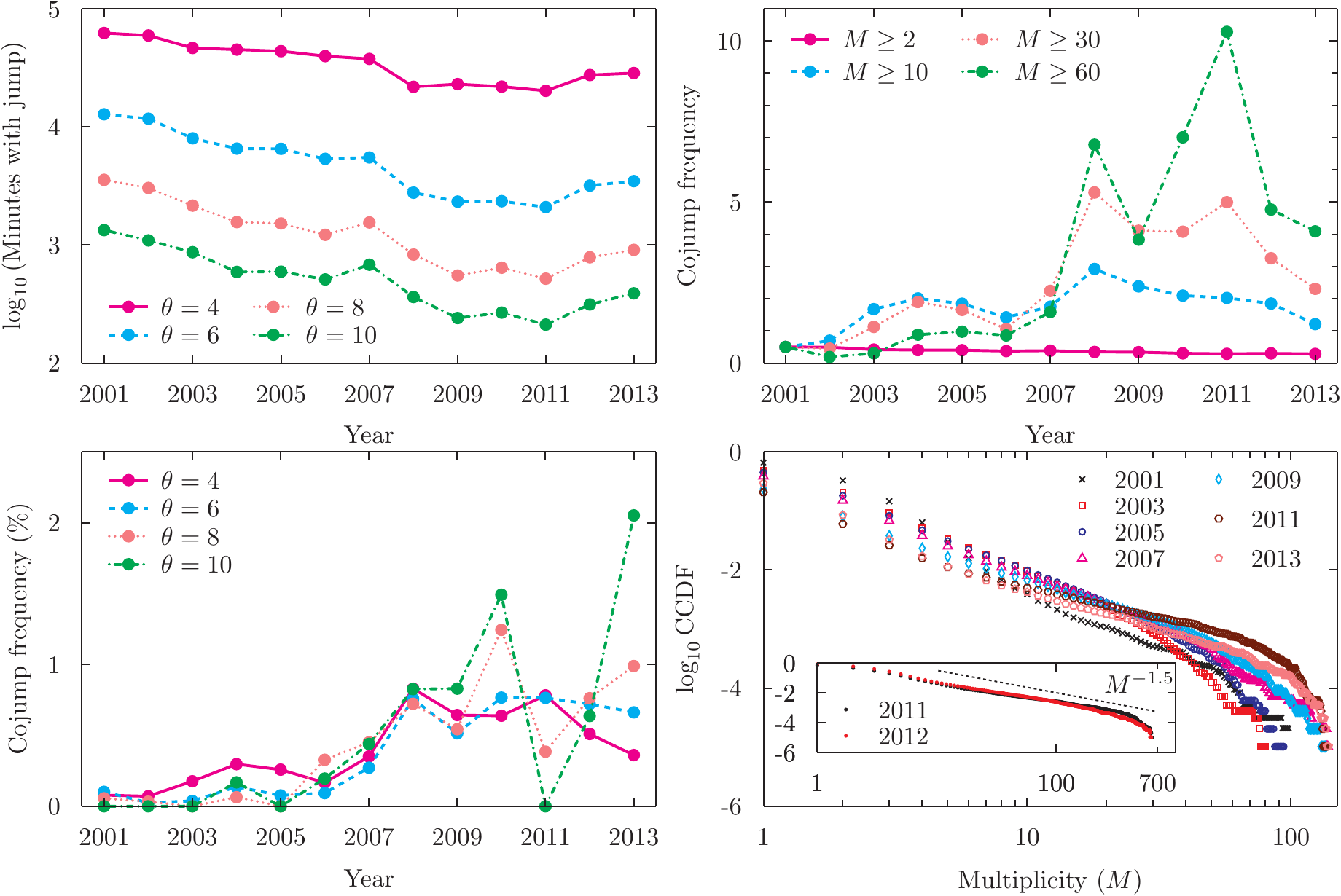}
  \caption{Top left panel: Semi-log plots of the total number of minutes where we detect at least one jump among the 140 selected assets of the Russell 3000 Index. Curves correspond to four different levels of the threshold parameter $\theta$. Top right panel: For $\theta=4$, yearly time evolution of the fraction of minutes with at least one event of multiplicity larger than or equal to 2, 10, 30, 60. All values are normalised by the corresponding 2001 values. Bottom left panel: Yearly evolution of the percentage fraction of cojumps with multiplicity at least equal to 30 for four different values of $\theta$. Bottom right panel: Log-log plots of the Complementary of the Cumulative Distribution Function of the cojump multiplicity for seven different years. The panel reports the empirical evidence for a portfolio of 140 stocks, while the inset details results of the same analysis conducted with 700 liquid assets from Russell 3000 during years 2011 and 2012.} \label{fig:multiplicity}
\end{figure*}
First, in the top left panel of Fig.~\ref{fig:multiplicity} we show the frequency of jumps per minute in each year, considering different values of $\theta$. We observe that for all $\theta$s the number of jumps has actually {\it decreased} over time. The different lines are quite parallel one to each other (especially for $\theta\ge 6$) indicating that the tails of the one minute return distribution remained quite stable. A completely different pattern emerges when we consider the dynamics of cojumps. The top right panel of Fig.~\ref{fig:multiplicity} shows the frequency of cojumps of different multiplicity (normalized to its value in 2001). While the frequency of cojumps with any multiplicity ($M\ge 2$) has slightly declined, the frequency of high multiplicity cojumps has become in recent years up to 10 times more frequent than its value in 2001. The result is essentially unchanged when fixing the minimal multiplicity (e.g.~$M\ge 30$) and computing the number of cojumps for different values of $\theta$ (bottom left panel of Fig.~\ref{fig:multiplicity}). Clearly larger fluctuations are observed for larger values of $M$. The increase of frequency of high multiplicity events is not due to the fact that markets have become faster. In the \emph{Support Information} we show the fraction of cojumps with $M \geq 30$ and $M \geq 60$ at $1,\ldots,5$ minutes. It is clear that the variability with the time window defining the event is much smaller than the secular variability of the events. In fact the fraction of 1-min cojumps with $M \geq 30$ in 2013 is significantly larger than the fraction of 5-min cojumps with $M \geq 30$ in 2001. The same is true for cojumps with $M \geq 60$. Therefore, the increase in synchronization is a genuine phenomenon, not explained by the increase in market speed. 

Finally, the bottom right panel of Fig.~\ref{fig:multiplicity} shows the distribution function of the cojump multiplicity for different years. Despite some variation is observed across the years, a clear power law tail behavior is evident. This means that the probability of systemic cojumps is quite large. Consistently with the observations above, the tail is thicker in recent years (even if in 2013 we observe a slightly thinner tail). It is important to notice that the bending of the distributions for large multiplicity is very likely due to the finite support of the distribution. Clearly for a set of $N$ stocks the multiplicity cannot be larger than $N$, thus the distribution function is zero at $M=N$. To show the role of the finite support, in the inset we show the multiplicity distribution function for a larger set of $700$ highly liquid assets. In this case the power law region extends for a wider range and close to $M=700$ we observe the expected bending of the function. The tail exponent of these distributions is close to $1.5$ (similarly to what observed in \cite{Joulin_etal:2008}).

In conclusion, at the beginning of 2000's individual jumps were more frequent and high frequency systemic instabilities, i.e.~high multiplicity jumps, were rare and mostly concentrated on macro-news announcements. In recent years, on the contrary, markets display often systemic cojumps and these are scattered across the trading day.

\subsection{Systemic cojumps and macroeconomic news}
\begin{figure}[t!]
  \centering
  \includegraphics[width=0.6\columnwidth]{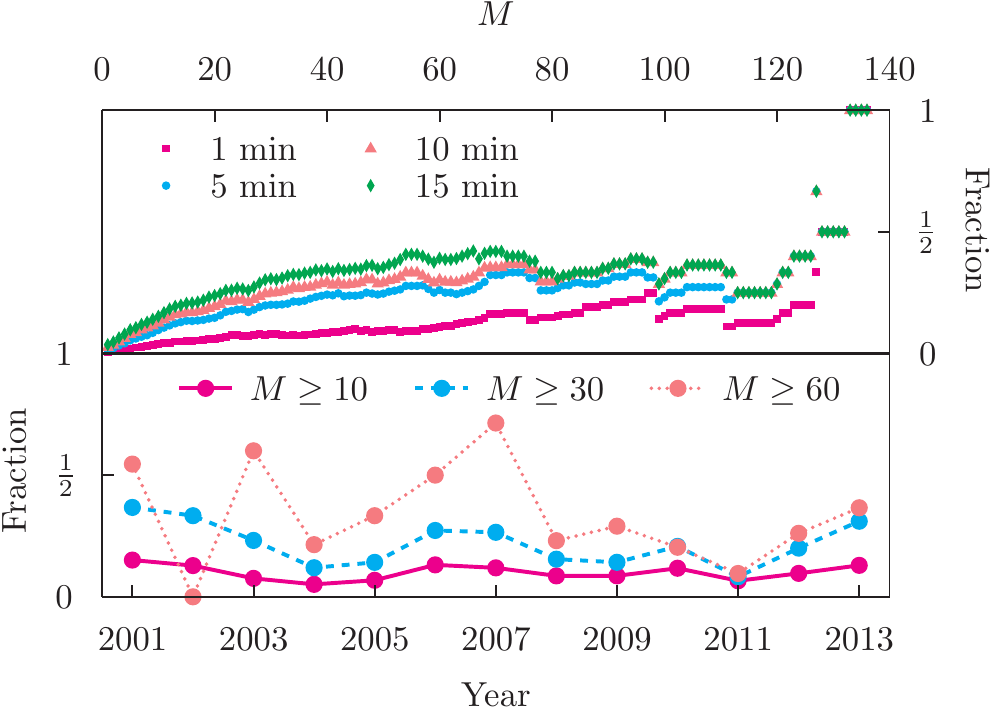}
  \caption{Top panel: Fraction of cojumps in 2012 with multiplicity larger than or equal to the value reported on the x axis for which a news occurred in the last 1, 5, 10, and 15 minutes. Bottom panel: Fraction of cojumps for different multiplicities $M$ for which we observe at least one news in a time window of five minutes preceding the jump event.} \label{fig:newsimpact}
\end{figure}
The second question is what fraction of these systemic cojumps has an exogenous or an endogenous origin. To answer this question we study how frequently a systemic cojump is preceded by a scheduled macroeconomic news. It is in fact unlikely that stock idiosyncratic news affect the whole market. We measure how frequently a systemic cojump with multiplicity larger than $M$ is preceded by a macronews in the last $\tau=1,5,10,15$ minutes. The top graph of Fig.~\ref{fig:newsimpact} shows that only 40\% of the high multiplicity cojumps are preceded by a macronews in the previous 15 minutes. Notice that the fractions of news-triggered systemic events in the 5, 10, and 15 minutes time windows are very close one to each other, indicating that if a macronews triggers a systemic cojump, this will typically happen within $5$ minutes from the news. 

For a historical perspective, the bottom graph of Fig.~\ref{fig:newsimpact} shows that the fraction of systemic cojumps triggered by macroeconomic news is quite constant across the years and, even for large $M$, clearly below 50\%. Thus our empirical analysis shows that a relevant portion of systemic cojumps is not associated with scheduled macroeconomic announcements. Idiosyncratic company-specific news may play a role, but plausibly only for those events which involve a very limited number of assets. For high multiplicity cojump events, endogenous mechanisms are likely to play a determinant role.

\section{Model} \label{sec_model}
\subsection{Hawkes process for multiplicity vector}

The empirical evidence of the previous section suggests that a large fraction of the dynamics of the systemic cojumps is unrelated to macro news and is likely endogenously generated. Moreover, as observed for example in the 2010 Flash Crash, market instabilities tend to propagate quickly to other assets, markets, or asset classes. Thus it is important to model the self- and cross-dependence of instabilities, considering both synchronous and lagged dependence, by studying whether and how systemic instabilities trigger other instabilities in the short run.

However the estimation of the interaction among a set of $140$ variables is extremely challenging and some sort of filtering is needed. A first step in this direction was taken in \cite{bormetti2015modelling} where we modeled the multivariate point process describing the jumps with a Hawkes factor model. Each stock is represented by a point process, each count being a jump. The coupling between the stocks is given by a one factor model structure, i.e. the intensity is the sum of the intensity of a factor and the intensity of an idiosyncratic term. Finally in order to capture the temporal clustering of events we assumed that both the factor and the idiosyncratic term follow a Hawkes process. 

As shown in \cite{bormetti2015modelling} this type of modeling is very effective (and parsimonious) in describing the pairwise properties of cojumps, i.e.~the probability that two stocks jump in the same time interval. However when considering cojumps of $M>2$ stocks, the model shows its weakness. An important indication is given by the distribution of multiplicities. It is possible to show that in the large $N$ limit, the factor model of \cite{bormetti2015modelling} predicts a multiplicity distribution with Gaussian tails, at odds with the power law behavior observed empirically in the bottom right panel of Fig.~\ref{fig:multiplicity}. Moreover the multiplicity of a systemic cojump is independent from the multiplicity of previous systemic cojumps, while the right panel of Fig.~\ref{fig:cojumps} shows clear temporal clusters of high multiplicity cojumps.

For these reasons, in this paper we propose a new modeling approach which preserves the parsimony and is able to overcome the problems of the model of \cite{bormetti2015modelling}. The idea is to   
model directly the vector of multiplicities, losing information on the identity of the cojumping stocks. 

Specifically, we consider an $N$-dimensional point process characterized by the vector of intensities $\bm{\lambda}_t$. An event in the $i$-component at time $t$ means that at this time a systemic cojump of multiplicity $i$ has occurred. Under this modeling assumption we know the total number of assets which have jumped, but we can no longer identify which companies among the $N$ possible ones have moved. To model the self- and cross-excitation of cojumps we use an $N$-dimensional Hawkes process with exponential kernels (see the \emph{Support Information} for the definition and the most relevant features). In general, the model depends parametrically on the baseline intensity vector $\bm{\mu}$, and on the $N\times N$ matrices $\alpha_{ij}$ and $\beta_{ij}$ of parameters characterizing the kernels. In order to reduce the dimension of the estimation problem from $N+2N^2$ to a more manageable number of unknowns, we proceed as follows. Since an important goal of our model is the ability to reproduce the empirical stationary distribution of the multiplicity vector, we assume $\bm{\mu}=\eta\mathbb{E}[\bm{\lambda}_t]$, where $0<\eta<1$, and $\mathbb{E}[\bm{\lambda}_t]$ proportional to the observed multiplicity frequencies. Interestingly, it is possible to show that $1-\eta$ is the spectral radius of the kernel matrix and therefore it measures the fraction of intensity explained by the self- and cross-excitation, while $\eta$ is the fraction explained by the baseline (exogenous) intensity.
We assume that all the parameters $\beta_{ij}$ which characterize the decay time of the self- and cross-excitations are equal to a constant value $\beta$.  Finally, we hypothesize that, for fixed $i=1,\ldots,N$, the largest intensity shock is ascribable to the self-exciting term $\alpha_{ii}$, while the cross-exciting effects as a function of the distance $|i-j|$ between multiplicities decrease hyperbolically with a tail exponent $\gamma$. This means that cojumps of a given multiplicity excite with higher probability cojumps with similar multiplicity. To sum up, the model is completely specified in terms of three parameters, $\eta$, $\beta$, and $\gamma$, and the empirical expected number of events with fixed multiplicity.

\subsection{Model results} \label{sec_model_results}
We apply the model to the dataset of 140 stocks in 2013. In order to calibrate and test the model we make use of two quantities, $f_\tau^{(1)}(M;J)$ and $f_\tau^{(2)}(M)$, defined in the \emph{Support Information}. The first one is the probability, conditional on the realization at time $t$ of an event with multiplicity at least $M$, of a cojump with multiplicity at least $J$ in the interval $(t,t+\tau]$. It measures how frequently a systemic cojump triggers other systemic cojumps in the short run. The second quantity is the average multiplicity of the cojumps inside a time interval of length $ \tau$ after a cojump of multiplicity larger than or equal to $M$. It therefore measures the typical cojump multiplicity triggered by a cojump of multiplicity at least $M$. We consider here the case $\tau=5$ minutes.

\begin{figure*}[t!]
  \centering
  \includegraphics[width=\textwidth]{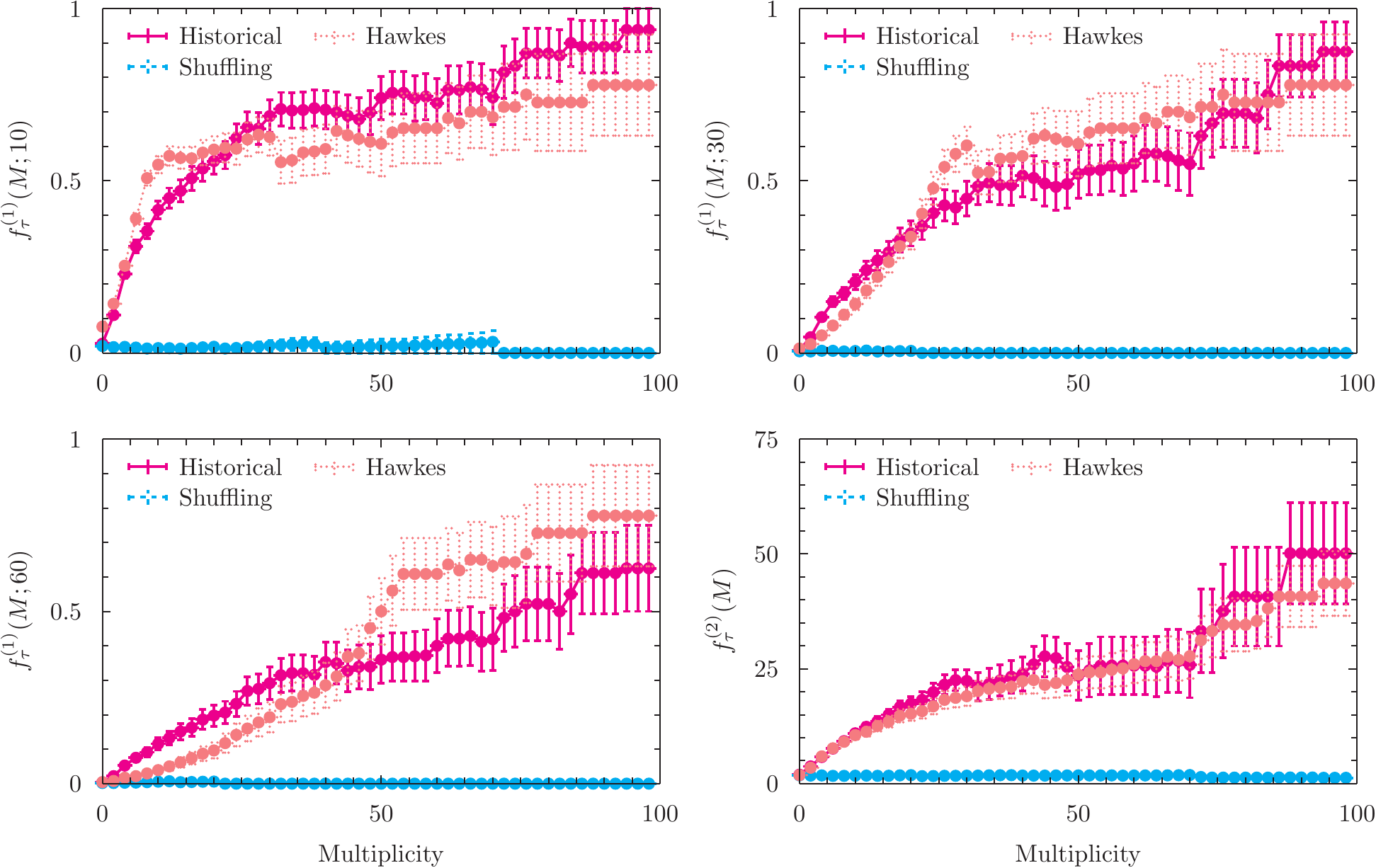}
  \caption{Top left panel: Probability that a cojump with multiplicity larger than or equal to 10 occurs in a $\tau=5$ minute interval following a cojump at time $t$ with multiplicity $M_t\geq M$. Plots are obtained from historical and simulated data. The error bars represent standard errors. Top right and bottom left panels: Threshold 10 replaced by 30 and 60, respectively. Bottom right panel: Expected amplitude of the cojumps in a $\tau=5$ minute interval following a cojump with multiplicity $M_t\geq M$.} \label{fig:indicator}
\end{figure*}
We use the $f_\tau^{(1)}(M;J)$ with $J=10$ and $f_\tau^{(2)}(M)$ to calibrate the model (see the \emph{Support Information} for details) and we test it on  $f_\tau^{(1)}(M;J)$ with $J=30$ and $J=60$. The  estimated parameters are $\eta = 0.15$, $\beta = 0.6$, $\gamma = 2.65$. Thus $85\%$ of the cojump activity is explained by the excitation mechanism and only $15\%$ is exogenous. The typical timescale of the memory is $1/\beta\simeq 1.67$ minutes and the relatively low value of $\gamma$ indicates a strong cross-excitation between different multiplicities.  As expected, the model effectively reproduces the stationary distribution of the multiplicities observed in empirical data (see Fig.~3 of the \emph{Support Information}). Fig.~\ref{fig:indicator} reports the quantities $f_\tau^{(1)}(M;J)$ and $f_\tau^{(2)}(M)$ in real and simulated data. The solid line corresponds to the empirical probabilities, the dotted line to the results from the Hawkes model, and as a benchmark case we also show the result of a shuffling experiment on the multiplicity time series (dashed line). It is evident that dropping the lagged correlations we obtain an unrealistic description of the multiplicity process. The Hawkes model, on the contrary, fits well the empirical data and therefore adequately describes the cross-excitation mechanism between systemic cojumps. Some discrepancies are observable for $J=60$, but the general shape of the curve and its level are well reproduced and the Hawkes model is a huge improvement with respect to the benchmark case. This evidence confirms that the larger is the value of the conditioning multiplicity the greater is the probability that in the subsequent minutes an event with large multiplicity happens.  

\section{Discussion} \label{sec:IV}
By investigating a portfolio of highly liquid stocks, our research enlightens a remarkable evidence: Since 2001 the total number of extreme events has remarkably diminished, but the number of occurrences where a sizable fraction of assets jump together has increased. This trend is more and more pronounced as we consider events of higher and higher multiplicity. This evidence is a clear mark that markets are nowadays more and more interconnected and a strong synchronization between jumps of different assets is present.

What are the factors responsible for the appearance of extreme movements? The cause can be either exogenous or endogenous. The former case is linked to the release of macro-economic news impacting the price dynamics, while the latter may result from unstable market conditions, such as a temporary lack of liquidity. Quite unexpectedly, only a minor fraction (up to 40\%) of the cojumps involving a large number of assets can be attributed to exogenous news. The remaining 60\% suggests that a more intriguing endogenous mechanism is taking place. Why has the synchronization among different assets increased through the recent years? We hypothesize that a major role is played by the dramatic increase of algorithmic trading. Thanks to the technological innovation, faster information processing is responsible for the more rapid propagation of large price movements through different assets. We also provide the evidence that highly systemic instabilities have the double effect of (i) increasing the probability that another systemic event takes place in the near future and (ii) increasing the degree of systemicity of short-term instabilities.

The low timescale of the memory of the exciting effects and the strong persistence of the cross-excitation among different multiplicities support the idea that, to achieve an accurate description of high frequency price dynamics, we should abandon conventional modeling assumptions. Coherently, we propose an innovative approach to the collective behavior of assets' prices based on the Hawkes description of the multiplicity process. Our model well describes the short term dynamics of systemic instabilities while preserving a remarkable parsimony in the number of parameters.  Thus, it provides a realistic description of the market behavior which is of prime importance from several perspectives, from trading to risk control, and market designing.  

\appendix

\section{Support Information: Data}

\subsection{Market data}
Data are provided by Kibot, \texttt{www.kibot.com}. We consider the thirteen years from 2001 to 2013 and for each year we select 140 highly liquid stocks in the Russell 3000 index. We exclude American Depositary Receipts, which are negotiable instruments representing ownership in non-US companies, since their dynamics is heavily influenced by their primary market and thus shows a peculiar intraday pattern. We use 1-minute closing price data during the regular US trading session, i.e.~from 9:30 a.m.~to 4:00 p.m. We discard early-closing days (typically, the eves of Independence Day, Thanksgiving and Christmas). Data are adjusted for splits and dividends.

Intraday returns are first filtered for the average intraday pattern, since price fluctuations are known to exhibit significant differences in absolute size depending on the time of the day, showing a typical U shape with larger movements at the beginning and at the end of the trading day. We perform this filtering in a standard way by dividing price returns by the intraday pattern, which is calculated as the average, over all days, of absolute returns rescaled by the daily volatility. Such normalised returns no longer possess any daily regularities and can thus be considered a unique time series with no periodic structure. For more details please refer to~\cite{bormetti2015modelling}. 

\subsection{News data}
\begin{sidewaystable}[t!]
  \caption{Number of news announcements, organized by year and news category.}
  \footnotesize
  \begin{tabular}{p{4cm}lrrrrrrrrrrrrrr}
    &     & 2001 & 2002 & 2003 & 2004 & 2005 & 2006 & 2007 & 2008 & 2009 & 2010 & 2011 & 2012 & 2013 & all years\\
    \hline
    ADP Employment Report                  & MEA\footnote{Merit Extra Attention according to the classification provided by Econoday, Inc.} &    0 &    1 &    0 &    0 &    0 &    0 &    0 &    0 &    0 &    0 &    0 &    0 &    0 &         1\\
    Beige Book                             & MEA &    8 &    8 &    8 &    7 &    8 &    8 &    8 &    8 &    8 &    8 &    8 &    8 &    8 &       103\\
    Business Inventories                   & MEA &    0 &    0 &    3 &    6 &    5 &   12 &   12 &   12 &   12 &   12 &   12 &   12 &   12 &       110\\
    Chairman Press Conference              & MMI\footnote{Market Moving Indicator according to the classification provided by Econoday, Inc.} &    0 &    0 &    0 &    0 &    0 &    0 &    0 &    0 &    0 &    0 &    3 &    5 &    4 &        12\\
    Chicago PMI                            & MEA &   12 &   12 &   12 &   12 &   12 &   12 &   12 &   12 &   12 &   12 &   12 &   12 &   11 &       155\\
    Construction Spending                  & MEA &   12 &   12 &   12 &   12 &   12 &   12 &   12 &   12 &   12 &   12 &   12 &   12 &   12 &       156\\
    Consumer Confidence                    & MEA &   12 &   12 &   12 &   12 &   12 &   12 &   12 &   12 &   12 &   13 &   12 &   12 &   11 &       156\\
    Consumer Sentiment                     & MEA &   12 &   23 &   24 &   23 &   24 &   24 &   24 &   24 &   24 &   24 &   24 &   24 &   24 &       298\\
    Dallas Fed Mfg Survey                  & MEA &    0 &    0 &    0 &    0 &    0 &    0 &    0 &    0 &    0 &    0 &    8 &   12 &   12 &        32\\
    Durable Goods Orders                   & MMI &    1 &    1 &    0 &    0 &    0 &    0 &    0 &    0 &    0 &    0 &    0 &    0 &    0 &         2\\
    EIA Petroleum Status Report            & MEA &    0 &    0 &    0 &   14 &   51 &   52 &   52 &   53 &   52 &   52 &   52 &   52 &   52 &       482\\
    Empire State Mfg Survey                & MEA &    0 &    0 &    0 &    0 &    0 &    1 &    0 &    0 &    0 &    0 &    0 &    0 &    0 &         1\\
    Existing Home Sales                    & MMI &   12 &   12 &   12 &   12 &   12 &   12 &   12 &   12 &   12 &   12 &   12 &   12 &   12 &       156\\
    Factory Orders                         & MEA &   12 &   12 &   12 &   12 &   12 &   12 &   12 &   12 &   12 &   12 &   12 &   12 &   12 &       156\\
    FOMC Forecasts                         & MMI &    0 &    0 &    0 &    0 &    0 &    0 &    0 &    0 &    0 &    0 &    0 &    5 &    4 &         9\\
    FOMC Meeting Announcement              & MMI &    8 &    8 &    8 &    8 &    8 &    8 &    8 &    8 &    8 &    8 &    8 &    8 &    8 &       104\\
    FOMC Minutes                           & MMI &    8 &    8 &    8 &    8 &    8 &    8 &    8 &    8 &    8 &    8 &    8 &    8 &    7 &       103\\
    Housing Market Index                   & MEA &    0 &    0 &    0 &    0 &    0 &   12 &   12 &   12 &   12 &   12 &   12 &   12 &   12 &        96\\
    ISM Mfg Index                          & MMI &   12 &   12 &   12 &   12 &   12 &   12 &   12 &   12 &   12 &   12 &   12 &   12 &   12 &       156\\
    ISM Non-Manufacturing Employment Index & MEA &    0 &    0 &    0 &    0 &    0 &    0 &    0 &   11 &   12 &    0 &    0 &    0 &    0 &        23\\
    Motor Vehicle Sales                    & MEA &    0 &   12 &    0 &    0 &    0 &    0 &    0 &    0 &    0 &    0 &    0 &    0 &    0 &        12\\
    New Home Sales                         & MMI &   13 &   12 &   12 &   12 &   12 &   12 &   12 &   12 &   12 &   12 &   12 &   12 &   12 &       157\\
    Pending Home Sales Index               & MEA &    0 &    0 &    0 &    0 &    0 &   12 &   12 &   12 &   12 &   13 &   12 &   12 &   12 &        97\\
    Personal Income and Outlays            & MMI &    1 &    0 &    0 &    0 &    0 &    0 &    0 &    0 &    0 &    0 &    0 &    0 &    0 &         1\\
    Philadelphia Fed Survey                & MMI &   12 &   12 &   12 &   12 &   12 &   12 &   12 &   12 &   12 &   12 &   12 &   12 &   12 &       156\\
    Retail Sales                           & MMI &    0 &    0 &    0 &    0 &    0 &    0 &    0 &    1 &    0 &    0 &    0 &    0 &    0 &         1\\
    Treasury Budget                        & MEA &   12 &   12 &   12 &   12 &   12 &   12 &   12 &   11 &   12 &   12 &   12 &   12 &   10 &       153\\
    \hline
    all categories                         &     &  147 &  169 &  159 &  174 &  212 &  245 &  244 &  256 &  256 &  246 &  255 &  266 &  259 &      2888\\
    \hline
  \end{tabular}
\label{table:si}
\end{sidewaystable}
The macronews dataset is provided by Econoday, Inc., \texttt{www.econoday.com}. Table \ref{table:si} shows the number of news announcements, organized by year and news category.

\section{Support Information: Dependence of systemic cojumps on time scale detection}
\begin{figure*}[t]
  \includegraphics[width=0.5\columnwidth]{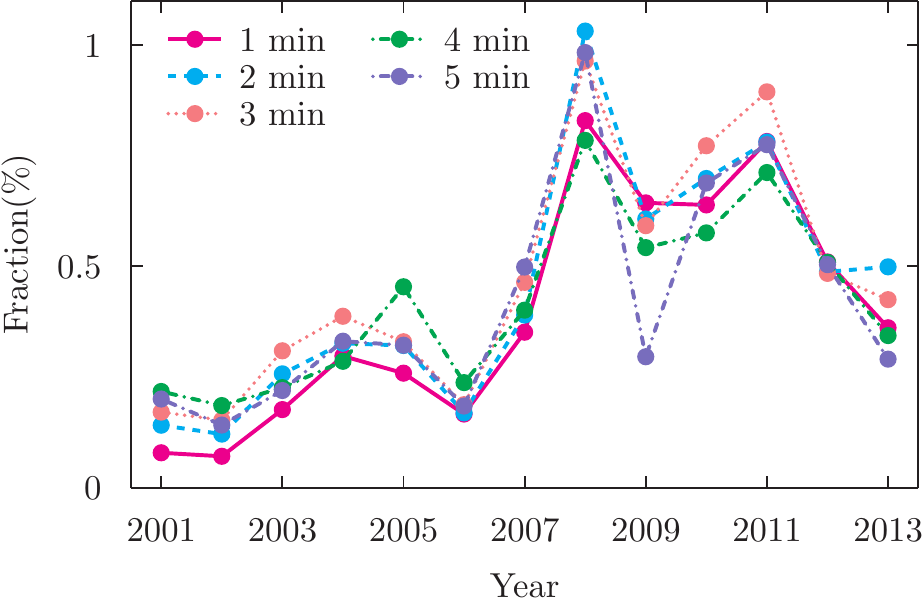}
  \includegraphics[width=0.5\columnwidth]{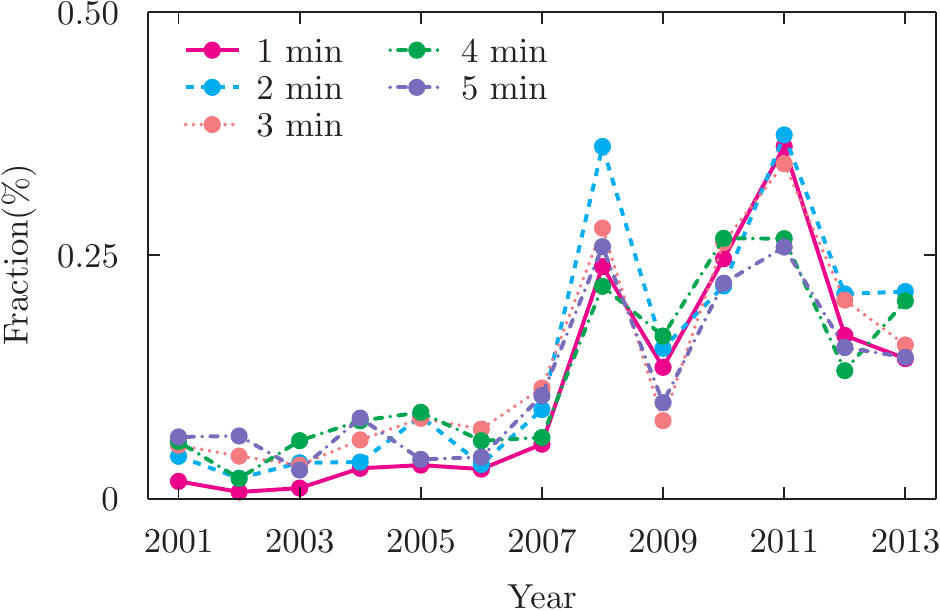}
  \caption{Yearly time evolution of the fraction of cojumps with multiplicity $M\ge 30$ (left) or $M\ge 60$ (right) over the total number of cojumps ($M\ge1$) for $\theta=4$ and different time horizons, namely $1,\ldots,5$ minutes.} \label{fig:timescale}
\end{figure*}
The paper mostly considers one minute (co)jumps. However one minute in 2013 is not equivalent to one minute in 2001 in terms of market activity. Hence it is important to test whether the increase in number of high multiplicity cojumps is due to the fact that in older years synchronization occurred on a time scale longer than one minute. To test this possibility we have repeated the analysis varying the time scale for jump detecttion from one to five minutes. Analyses on the dynamics of cross-correlation between stocks data suggested us that the time scale over which stocks become correlated has decreased by a factor approximately equal to five from 2001 to 2013.

Fig.~\ref{fig:timescale} shows the yearly time evolution of the fraction of cojumps with multiplicity $M \ge 30$ (left) or $M \ge 60$ (right) over the total number of cojumps ($M \ge 1$) for $\theta = 4$ and different time scales, namely $1,\ldots,5$ minutes. Except for the first two years, no clear sorting of this fraction with the time scale is detectable, while the global secular trend has a much larger variability. This is particularly evident for the $M \ge 60$ case. Hence the number of high multiplicity one minute cojumps in 2013 is much higher than the number of high multiplicity five minute cojumps in 2001, indicating that the increased speed of market activity is a minor cause of the increase of high multiplicity systemic cojumps in recent years.

\section{Support Information: Model}

In the paper we model the point process describing the cojumps of $k$ stocks (independently from their identity) as the $k$-th component of a multivariate Hawkes process. These processes were introduced in the early Seventies~\cite{Hawkes:1971}, and have been widely employed to model earthquake data~\cite{Vere-Jones:1970,Vere-Jones_Ozaki:1982,Ogata:1988}. For a complete overview of the properties of Hawkes processes please refer to~\cite{Daley_Vere-Jones:2003,Bauwens_Hautsch:2009}, while for a review of their recent applications in a financial context see~\cite{Bacry_etal:2015}. Here we detail how we build and estimate the model. 

\subsection{Multivariate Hawkes point processes}

An $N$-dimensional Hawkes process is a point process characterized by the vector of intensities $\bm{\lambda}_t:=\left(\lambda_t^1,\ldots,\lambda_t^N\right)^\intercal$, where the $i$-type intensity satisfies the relation
\bes
\lambda_t^i = \mu_t^i + \sum_{j=1}^N \sum_{t_k^j < t} \nu_j^i(t-t_k^j)\,,
\ees
where $\mu_t^i$ and $\nu_j^i$ are positive deterministic functions for all $i,j=1,\ldots,N$. The set $\left\{t_k^j\right\}$ corresponds to the random sequence of increasing events associated with the $j$-component of the $N$-dimensional point process. If $\mu_t^i = \mu^i$ is a constant and the kernel function $\nu_j^i$ reduces identically to zero, then the Hawkes point process describing the $i$-component reduces to a Poisson process with constant intensity $\mu^i$. On the contrary, if the kernel is positive, each time an event occurs for any component of the multidimensional process, the intensity $\lambda_t^i$ increases by a positive amount. 

\subsection{Choice of the parametrization}

As in most high-dimensional problems, the estimation of multivariate Hawkes processes is problematic because of the large number of parameters. In order to overcome the curse of dimensionality problem, in this paper we choose a quite rigid parametrization of the kernel matrix, reducing significantly the number of free parameters. We also propose a method to estimate the model on data.

First of all, we assume that the vector $\bm{\mu}:=\left(\mu_t^1,\ldots,\mu_t^N\right)^\intercal$ does not depend on time. Second, we consider the most common parametrization of the kernel in terms of exponential functions
\bes
\nu_j^i(t-t_k^j):= \alpha_{ij}\ue^{-\beta_{ij}(t-t_k^j)}\,,
\ees
with $\alpha_{ij}> 0$ and $\beta_{ij}>0$ for all $i, j$. The parameter $\alpha_{ij}$ fixes the scale of the intensity process $\lambda^i$ and provides the deterministic amount by which the $j$-type event at $t_k^j$ shocks the intensity of the $i$-type process. The parameter $\beta_{ij}$ describes the inverse of the time needed by the process $i$ to lose memory of a count of process $j$.

The process is stationary if the spectral radius (i.e. the absolute value of the largest eigenvalue) of the matrix $\Gamma$ of elements
\bes
\Gamma_{ij}=\frac{\alpha_{ij}}{\beta_{ij}}
\ees
is strictly smaller than one. In this case the unconditional expected intensities of the process reads
\be
\label{eq:expected}
\mathbb{E}\left[\bm{\lambda}_t\right]=(\mathrm{I}_N-\Gamma)^{-1}\bm{\mu}\,, 
\ee
where $\mathrm{I}_N$ is the $N$-dimensional identity matrix. 

We make the following further assumptions:
\begin{itemize}
\item We assume that all the $\beta_{ij}$ are equal to a constant value $\beta>0$. This means that there is only one time scale characterizing the decay of the kernels.
\item We impose the condition that $\bm{\mu}=\eta \mathbb{E}\left[\bm{\lambda}_t\right]$, with $0<\eta<1$. This means that the distribution of multiplicity in the observed process is the same as the distribution of the multiplicity in the baseline (or ancestor) process. In other words, the cross-excitation between the different components of the Hawkes process does not change the unconditional law of multiplicity. Notice that this assumption implies that 
\bes
\Gamma\mathbb{E}\left[\bm{\lambda}_t\right]=(1-\eta)\mathbb{E}\left[\bm{\lambda}_t\right]\,, 
\ees
i.e. $\mathbb{E}\left[\bm{\lambda}_t\right]$ (or $\bm{\mu}$) is the eigenvector of $\Gamma$ with eigenvalue $1-\eta$ . 
\item The generic matrix element $\Gamma_{ij}$ describing the intensity of the excitation of variable $j$ on variable $i$ is the product of a term $D_{ii}$ which depends on the excited variable and a term $\sigma(|i-j|)$ which depends on the absolute difference of the two multiplicities. Therefore we can rewrite  $\Gamma= D\Sigma$, where $D$ is a diagonal matrix of elements
\bes
D_{ii}:=\frac{(1-\eta)\mu^i}{\sum_{j=1}^N \mu^j\sigma(|i-j|)},
\ees
 and $\Sigma_{ij}=\sigma(|i-j|)$.
\item Finally, we parametrize the matrix $\Sigma$ as
\bes
\Sigma_{ij}=\sigma(|i-j|)=(|i-j| + 1)^{-\gamma}
\ees
This hyperbolic decay is chosen to model with only one parameter $\gamma$ the strong cross-excitation between two very different multiplicities.
\end{itemize}
The model is therefore parametrized by the vector $\bm{\mu}$ and the three parameters $\eta$, $\gamma$, and $\beta$.

Before presenting the estimation procedure, we discuss some properties of the model. As all the entries of $\Gamma$ are strictly positive, the Perron-Frobenius Theorem applies. Then, there exists only one eigenvector with all strictly positive components, and the associated eigenvalue is the spectral radius. Since $\mathbb{E}\left[\lambda_t^i\right]>0$ for all $i=1,\ldots,N$, we conclude that the spectral radius is $1-\eta$. Incidentally, we notice that all the eigenvalues of $\Gamma$ are real. This property readily follows from observing that $\Gamma$ is the product of two symmetric matrices, and $D$ is diagonal and positive definite. Indeed, denoting with $\sqrt{D}$ the square root of the matrix $D$, $\Gamma$ is \textit{similar} to $\sqrt{D}^{-1} D \Sigma \sqrt{D}$, which is by construction symmetric. Moreover, if $\Gamma$ is diagonal dominant, i.e. if $|\Gamma_{ii}| > \sum_{j\neq i} |\Gamma{ij}|$ for $i=1,\ldots,N$, the eigenvalues are also strictly positive.  

\subsection{Estimation of the model parameters}
A rigorous estimation of our model's parameters through likelihood maximization poses several computational problems. We instead propose a heuristic and robust calibration procedure based on moments. In particular we consider the following two conditional expectations, whose values on real and simulated data are graphed in Fig.~4 of the main article:
\begin{equation} \label{eq:CE1}
	f_\tau^{(1)}(M;J):=\mathbb{P} \Big[ \exists t' \in \, (t,t + \tau] \ \text{s.t.} \ M_{t'} \ge J \Big| M_t \ge M \Big] \, ,
\end{equation}
\begin{equation} \label{eq:CE2}
	f_\tau^{(2)}(M):=\mathbb{E} \Big[ M_{t'} \Big| M_t \ge M , \, \exists t' \in \, (t,t + \tau] \ \text{s.t.} \ M_{t'} > 0 \Big] \, .
\end{equation}
The first quantity, $f_\tau^{(1)}(M;J)$, is the probability of observing a systemic event with multiplicity at least $J$ inside a time interval of length $\tau$ after a cojump of multiplicity $M_t$ larger than or equal to $M$. It therefore measures the probability that a cojump of multiplicity at least $M$ triggers a systemic cojump ($J$ fixes the threshold for a systemic cojump). The second quantity, $f_\tau^{(2)}(M)$, is the average multiplicity of the cojumps inside a time interval of length $ \tau$ after a cojump of multiplicity $M_t$ larger than or equal to $M$. It therefore measures the typical cojump multiplicity triggered by a cojump of multiplicity at least $M$.

We use $f_\tau^{(1)}(M;J)$ (for fixed $J$ and $\tau$) and $f_\tau^{(2)}(M)$ (for fixed $\tau$) to estimate via a weighted least squares approach the three model parameters $\eta$, $\gamma$, and $\beta$. Since we are not able to compute analytically the moments of $f_\tau^{(1)}(M;J)$ and $f_\tau^{(2)}(M)$ from the model, we perform Monte Carlo simulations with fixed parameters.  Specifically, given a multiplicity $M$, let the data and the model conditional expectations of any of the quantity in Eq. [\ref{eq:CE1}] and [\ref{eq:CE2}] be represented by their average values $a_\text{d} (M)$, $a_\text{m} (M)$ and standard errors $\delta_\text{d} (M)$, $\delta_\text{m} (M)$. Then, for the expectation $f_\tau^{(i)}$ ($i=1,2$) we construct the loss function
\begin{equation} \label{eq:chi2}
	\chi^2_{(i)} = \sum_{M \in S} \frac{(a_\text{d} - a_\text{m})^2}{\delta_\text{d}^2 + \delta_\text{m}^2} \, ,
\end{equation}
where the sum is taken over a set of multiplicities $S$. We then construct the total loss function $\chi_{(1)}^2 + 0.5 \chi_{(2)}^2$ and we search for the model parameters which minimize the loss function. Given the small number of parameters we explore a large region of the three-dimensional space of parameters on a 0.05-spaced grid.

\subsection{Results for the investigated dataset}
\begin{figure*}[t!]
  \includegraphics[width=\columnwidth]{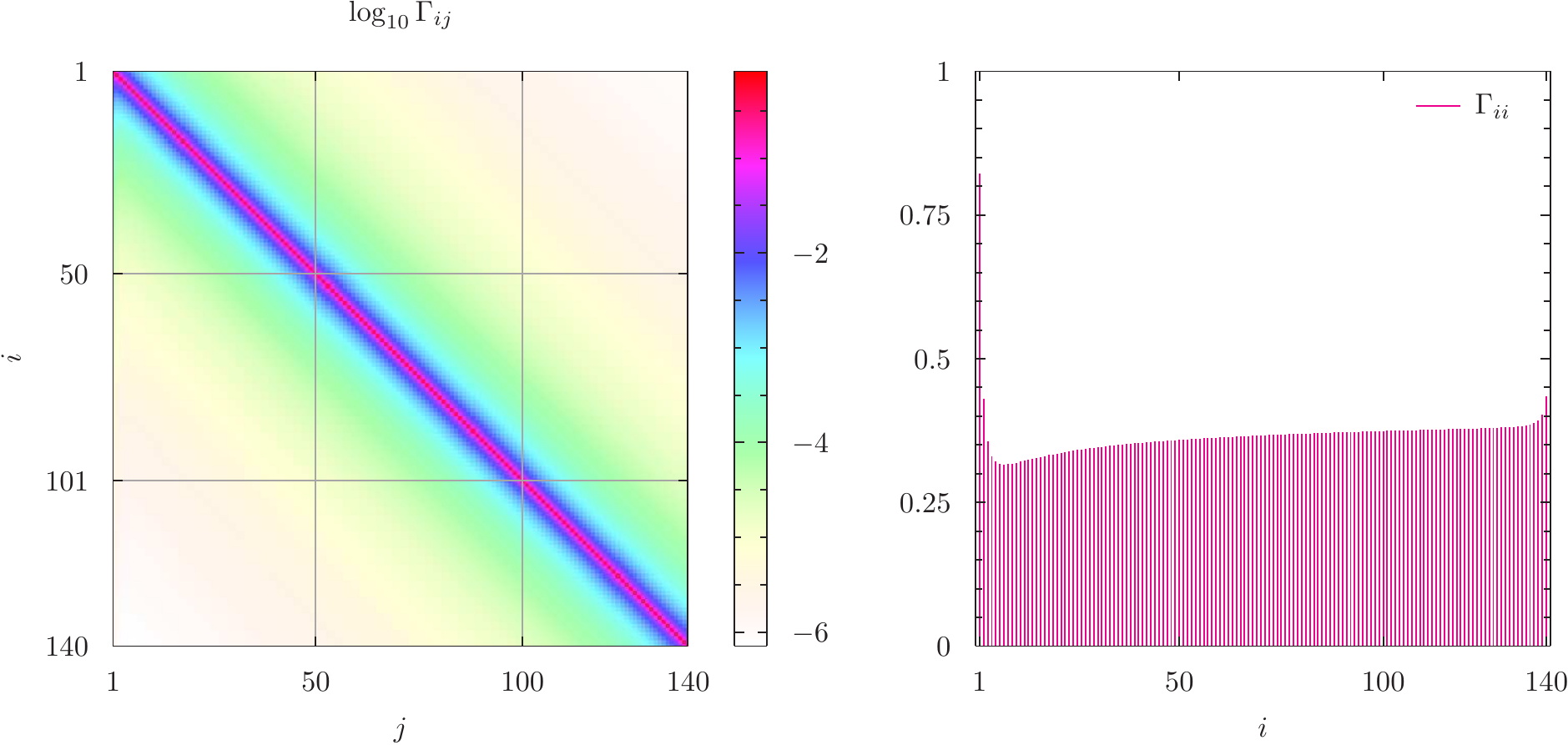}
  \caption{Left panel: Logarithmic entries of the matrix $\Gamma_{ij}:=\alpha_{ij}/\beta_{ij}$ for $\beta_{ij}=\beta=0.6$ for all $i,j=1,\ldots,140$, $\eta=0.15$, and $\gamma=2.65$. Right panel: Linear plot of the diagonal entries of $\Gamma$ as a function of the multiplicity $i$.} \label{fig:gamma}
\end{figure*}
As an example of the estimation procedure and to discuss the properties of the fitted model, we consider in detail the case of $N=140$ highly liquid assets of the Russell 3000 Index in 2013. The same set is used also in Fig.~4 of the main text. We fix $J = 10$ in Eq. [\ref{eq:CE1}], $\tau = 5$ in Eq. [\ref{eq:CE1}] and [\ref{eq:CE2}], $S = \{ 5, 10, 15, \ldots, 65, 70 \}$ and look for the parameters that minimise the total loss function. Following this approach, we find a clear minimum corresponding to the values $\eta = 0.15$, $\beta = 0.6$, $\gamma = 2.65$.
\begin{figure}[t]
  \centering
  \includegraphics[width=0.6\columnwidth]{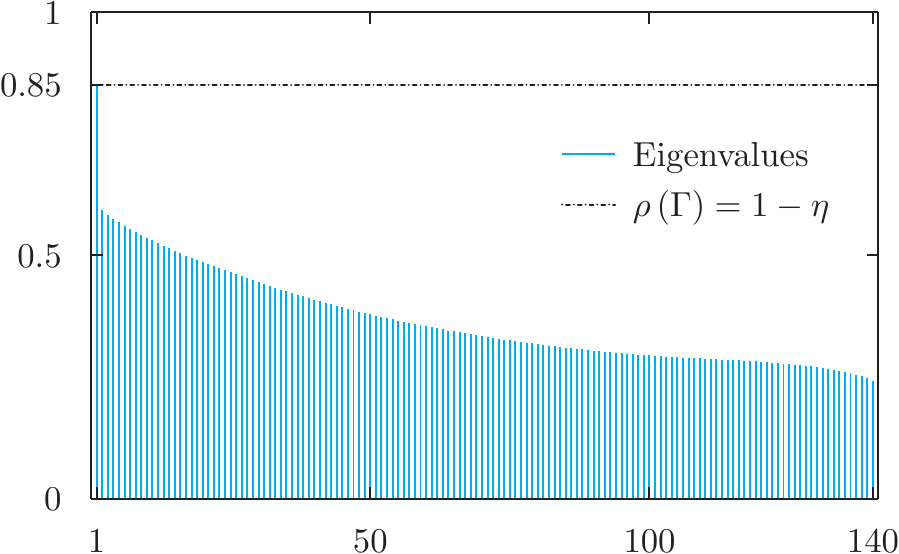}
  \caption{Eigenvalue spectrum of the matrix $\Gamma$. The spectral radius $\rho(\Gamma)$ corresponds to $1-\eta$. Since $\eta=0.15$, and more generally for $0<\eta<1$, the multidimensional Hawkes process describing the stochastic evolution of the multiplicity remains stationary. For the chosen parameter values, we verified numerically that $\Gamma$ satisfies the diagonal dominant condition and so all its eigenvalues are strictly positive.} \label{fig:eigenval}
\end{figure}

\begin{figure}[t!]
  \centering
  \includegraphics[width=0.6\columnwidth]{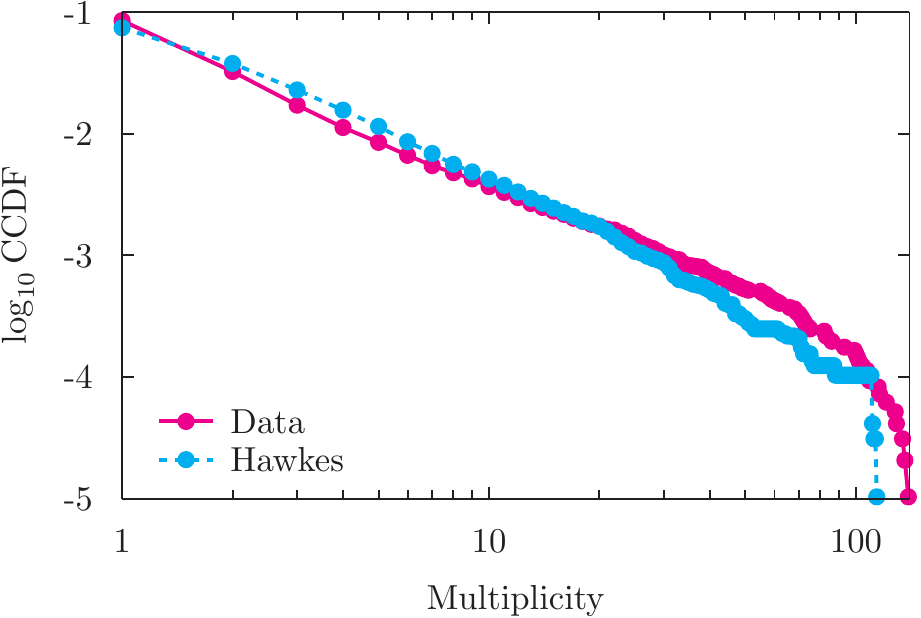}
  \caption{Log-log plot of the Complementary of the Cumulative Distribution Function of the cojump multiplicities. The bold line corresponds to the empirical distribution measured from the Russell 3000 data sample, 140 assets, during year 2013. The dashed line is the distribution obtained from a simulation of the multidimensional Hawkes process. The total number of minutes drawn from the simulation coincides with the length of the empirical time series and is equal to 96861.} \label{fig:CCDF}
\end{figure}

The left panel of Fig.~\ref{fig:gamma} reports the logarithmic value of $140\times 140$ entries of the $\Gamma$ matrix.  Coherently with the definitions given above, $\Gamma_{ij}$ for fixed $i$, is the impact of past events with multiplicity $j$ on the multiplicity $i$. The largest value corresponds to the diagonal term $\Gamma_{ii}=D_{ii}$ and quantifies the shock of the intensity due to a self-exciting effect. Then, moving away from the $\Gamma_{ii}$, the kernel matrix decreases symmetrically along the row according to a hyperbolic scaling with tail index $\gamma=2.65$. The parameter $\eta$ rescales the level of the main diagonal of the matrix $\Gamma$, reported in the right panel of Fig.~\ref{fig:gamma}, and determines the degree of stationarity of the process. In Fig.~\ref{fig:eigenval} we plot the complete spectrum of the matrix $\Gamma$. As expected, the largest value corresponds to $1-\eta=0.85$, while the positive definiteness of all the eigenvalues follows from the evidence, verified numerically, that the matrix is diagonal dominant. More specifically, for the chosen values of $\eta$, $\beta$, and $\gamma$ the matrix $\Gamma$ is determined uniquely through the specification of the vector of expected intensities, $\mathbb{E}\left[\bm{\lambda}_t\right]$. In our numerical experiment we replace the vector of expected intensities multiplied by the length of the time series, i.e. 96,861, with the empirical frequencies observed for the 140 assets from the Russell 3000 Index in 2013. Fig.~\ref{fig:CCDF} conveys this information in terms of the Complementary of the Cumulative Distribution Function of the cojump multiplicities associated with the empirical data (bold line). We also report the same quantity measured from a synthetic time series corresponding to a Monte Carlo simulation of the 140-dimensional Hawkes process (dashed line).

\section*{Acknowledgments}
FL acknowledges partial support by the grant SNS13LILLB ''Systemic risk in financial markets across time scales". Authors thanks E. Bacry, V. Filimonov, and M. Rambaldi for useful discussions.
The opinions expressed here are solely those of the authors and do not represent in any way those of their employers.


\begin{thebibliography}{40}
    \expandafter\ifx\csname url\endcsname\relax
    \def\url#1{\texttt{#1}}\fi
    \expandafter\ifx\csname urlprefix\endcsname\relax\def\urlprefix{URL }\fi
    \providecommand{\bibinfo}[2]{#2}
    \providecommand{\eprint}[2][]{\url{#2}}

  \bibitem{lewis2014flash}
    \bibinfo{author}{Lewis, M.}
    \newblock \emph{\bibinfo{title}{Flash boys: a Wall Street revolt}}
    (\bibinfo{publisher}{WW Norton \& Company}, \bibinfo{year}{2014}).

  \bibitem{gomber2011high}
    \bibinfo{author}{Gomber, P.}, \bibinfo{author}{Arndt, B.},
    \bibinfo{author}{Lutat, M.} \& \bibinfo{author}{Uhle, T.}
    \newblock \bibinfo{title}{High-frequency trading}.
    \newblock \emph{\bibinfo{journal}{Available at SSRN 1858626}}
    (\bibinfo{year}{2011}).

  \bibitem{macintosh2013high}
    \bibinfo{author}{MacIntosh, J.~G.}
    \newblock \bibinfo{title}{High frequency traders: Angels or devils?}
    \newblock \emph{\bibinfo{journal}{CD Howe Institute Commentary}}
    \textbf{\bibinfo{volume}{391}} (\bibinfo{year}{2013}).

  \bibitem{gerig2013high}
    \bibinfo{author}{Gerig, A.}
    \newblock \bibinfo{title}{High-frequency trading synchronizes prices in
    financial markets}.
    \newblock \emph{\bibinfo{journal}{Available at SSRN 2173247}}
    (\bibinfo{year}{2013}).

  \bibitem{sec2010}
    \bibinfo{title}{Findings regarding the market events of {M}ay 6, 2010. report
    of the staffs of the {CFTC} and {SEC} to the joint advisory committee on
  emerging regulatory issues}.
  \newblock \emph{\bibinfo{journal}{Available online at:
  www.sec.gov/news/studies/2010/marketevents-report.pdf}}
  (\bibinfo{year}{2010}).

\bibitem{kirilenko2014flash}
  \bibinfo{author}{Kirilenko, A.~A.}, \bibinfo{author}{Kyle, A.~S.},
  \bibinfo{author}{Samadi, M.} \& \bibinfo{author}{Tuzun, T.}
  \newblock \bibinfo{title}{The {F}lash {C}rash: The impact of high frequency
trading on an electronic market}.
\newblock \emph{\bibinfo{journal}{Available at SSRN 1686004}}
(\bibinfo{year}{2014}).

\bibitem{johnson2013abrupt}
  \bibinfo{author}{Johnson, N.} \emph{et~al.}
  \newblock \bibinfo{title}{Abrupt rise of new machine ecology beyond human
  response time}.
  \newblock \emph{\bibinfo{journal}{Scientific reports}}
  \textbf{\bibinfo{volume}{3}} (\bibinfo{year}{2013}).

\bibitem{golub2012high}
  \bibinfo{author}{Golub, A.}, \bibinfo{author}{Keane, J.} \&
  \bibinfo{author}{Poon, S.-H.}
  \newblock \bibinfo{title}{High frequency trading and mini flash crashes}.
  \newblock \emph{\bibinfo{journal}{arXiv preprint arXiv:1211.6667}}
  (\bibinfo{year}{2012}).

\bibitem{bormetti2015modelling}
  \bibinfo{author}{Bormetti, G.} \emph{et~al.}
  \newblock \bibinfo{title}{Modelling systemic price cojumps with {H}awkes factor
models}.
\newblock \emph{\bibinfo{journal}{Quantitative Finance}}
(\bibinfo{year}{2015}).
\newblock \bibinfo{note}{DOI: 10.1080/14697688.2014.996586}.

\bibitem{andersen2007no}
  \bibinfo{author}{Andersen, T.~G.}, \bibinfo{author}{Bollerslev, T.} \&
  \bibinfo{author}{Dobrev, D.}
  \newblock \bibinfo{title}{No-arbitrage semi-martingale restrictions for
    continuous-time volatility models subject to leverage effects, jumps and iid
  noise: Theory and testable distributional implications}.
  \newblock \emph{\bibinfo{journal}{Journal of Econometrics}}
  \textbf{\bibinfo{volume}{138}}, \bibinfo{pages}{125--180}
  (\bibinfo{year}{2007}).

\bibitem{lee2008jumps}
  \bibinfo{author}{Lee, S.~S.} \& \bibinfo{author}{Mykland, P.~A.}
  \newblock \bibinfo{title}{Jumps in financial markets: A new nonparametric test
  and jump dynamics}.
  \newblock \emph{\bibinfo{journal}{Review of Financial studies}}
  \textbf{\bibinfo{volume}{21}}, \bibinfo{pages}{2535--2563}
  (\bibinfo{year}{2008}).

\bibitem{andersen2010continuous}
  \bibinfo{author}{Andersen, T.~G.}, \bibinfo{author}{Bollerslev, T.},
  \bibinfo{author}{Frederiksen, P.} \& \bibinfo{author}{{\O}rregaard~Nielsen,
M.}
\newblock \bibinfo{title}{Continuous-time models, realized volatilities, and
testable distributional implications for daily stock returns}.
\newblock \emph{\bibinfo{journal}{Journal of Applied Econometrics}}
\textbf{\bibinfo{volume}{25}}, \bibinfo{pages}{233--261}
(\bibinfo{year}{2010}).

\bibitem{dumitru2012identifying}
  \bibinfo{author}{Dumitru, A.-M.} \& \bibinfo{author}{Urga, G.}
  \newblock \bibinfo{title}{Identifying jumps in financial assets: a comparison
  between nonparametric jump tests}.
  \newblock \emph{\bibinfo{journal}{Journal of Business \& Economic Statistics}}
  \textbf{\bibinfo{volume}{30}}, \bibinfo{pages}{242--255}
  (\bibinfo{year}{2012}).

\bibitem{bollerslev2013jump}
  \bibinfo{author}{Bollerslev, T.}, \bibinfo{author}{Todorov, V.} \&
  \bibinfo{author}{Li, S.~Z.}
  \newblock \bibinfo{title}{Jump tails, extreme dependencies, and the
  distribution of stock returns}.
  \newblock \emph{\bibinfo{journal}{Journal of Econometrics}}
  \textbf{\bibinfo{volume}{172}}, \bibinfo{pages}{307--324}
  (\bibinfo{year}{2013}).

\bibitem{gilder2014cojumps}
  \bibinfo{author}{Gilder, D.}, \bibinfo{author}{Shackleton, M.~B.} \&
  \bibinfo{author}{Taylor, S.~J.}
  \newblock \bibinfo{title}{Cojumps in stock prices: Empirical evidence}.
  \newblock \emph{\bibinfo{journal}{Journal of Banking \& Finance}}
  \textbf{\bibinfo{volume}{40}}, \bibinfo{pages}{443--459}
  (\bibinfo{year}{2014}).

\bibitem{caporin2014multi}
  \bibinfo{author}{Caporin, M.}, \bibinfo{author}{Kolokolov, A.} \&
  \bibinfo{author}{Ren{\`o}, R.}
  \newblock \bibinfo{title}{Multi-jumps}.
  \newblock \emph{\bibinfo{journal}{Available at SSRN 2488603}}
  (\bibinfo{year}{2014}).

\bibitem{petersen2010quantitative}
  \bibinfo{author}{Petersen, A.~M.}, \bibinfo{author}{Wang, F.},
  \bibinfo{author}{Havlin, S.} \& \bibinfo{author}{Stanley, H.~E.}
  \newblock \bibinfo{title}{Quantitative law describing market dynamics before
  and after interest-rate change}.
  \newblock \emph{\bibinfo{journal}{Physical Review E}}
  \textbf{\bibinfo{volume}{81}}, \bibinfo{pages}{066121}
  (\bibinfo{year}{2010}).

\bibitem{petersen2010market}
  \bibinfo{author}{Petersen, A.~M.}, \bibinfo{author}{Wang, F.},
  \bibinfo{author}{Havlin, S.} \& \bibinfo{author}{Stanley, H.~E.}
  \newblock \bibinfo{title}{Market dynamics immediately before and after
    financial shocks: Quantifying the {Omori}, productivity, and {Bath} laws}.
    \newblock \emph{\bibinfo{journal}{Physical Review E}}
    \textbf{\bibinfo{volume}{82}}, \bibinfo{pages}{036114}
    (\bibinfo{year}{2010}).

  \bibitem{lee2011jumps}
    \bibinfo{author}{Lee, S.~S.}
    \newblock \bibinfo{title}{Jumps and information flow in financial markets}.
    \newblock \emph{\bibinfo{journal}{Review of Financial Studies}}
    \bibinfo{pages}{hhr084} (\bibinfo{year}{2011}).

  \bibitem{bajgrowicz2013jumps}
    \bibinfo{author}{Bajgrowicz, P.}, \bibinfo{author}{Scaillet, O.} \&
    \bibinfo{author}{Treccani, A.}
    \newblock \bibinfo{title}{Jumps in high-frequency data: Spurious detections,
    dynamics, and news}.
    \newblock \emph{\bibinfo{journal}{Swiss Finance Institute Research Paper}}
    (\bibinfo{year}{2013}).

  \bibitem{Hawkes:1971}
    \bibinfo{author}{Hawkes, A.~G.}
    \newblock \bibinfo{title}{Spectra of some self-exciting and mutually exciting
    point processes}.
    \newblock \emph{\bibinfo{journal}{Biometrika}} \textbf{\bibinfo{volume}{58}},
    \bibinfo{pages}{83--90} (\bibinfo{year}{1971}).

  \bibitem{bowsher2007modelling}
    \bibinfo{author}{Bowsher, C.~G.}
    \newblock \bibinfo{title}{Modelling security market events in continuous time:
    Intensity based, multivariate point process models}.
    \newblock \emph{\bibinfo{journal}{Journal of Econometrics}}
    \textbf{\bibinfo{volume}{141}}, \bibinfo{pages}{876--912}
    (\bibinfo{year}{2007}).

  \bibitem{bauwens2009modelling}
    \bibinfo{author}{Bauwens, L.} \& \bibinfo{author}{Hautsch, N.}
    \newblock \emph{\bibinfo{title}{Modelling financial high frequency data using
    point processes}} (\bibinfo{publisher}{Springer}, \bibinfo{year}{2009}).

  \bibitem{Muni-Toke:2011}
    \bibinfo{author}{Muni~Toke, I.}
    \newblock \bibinfo{title}{``{M}arket making'' in an order book model and its
  impact on the spread}.
  \newblock In \bibinfo{editor}{Abergel, F.}, \bibinfo{editor}{Chakrabarti,
  B.~K.}, \bibinfo{editor}{Chakraborti, A.} \& \bibinfo{editor}{Mitra, M.}
  (eds.) \emph{\bibinfo{booktitle}{Econophysics of Order-Driven Markets}},
  \bibinfo{pages}{49--64} (\bibinfo{publisher}{Springer-Verlag, Milan},
  \bibinfo{year}{2011}).

\bibitem{Muni-Toke_Pomponio:2012}
  \bibinfo{author}{Muni~Toke, I.} \& \bibinfo{author}{Pomponio, F.}
  \newblock \bibinfo{title}{Modelling trades-through in a limit order book using
    {H}awkes processes}.
    \newblock \emph{\bibinfo{journal}{Economics: The Open-Access, Open-Assessment
    E-Journal}} \textbf{\bibinfo{volume}{6}}, \bibinfo{pages}{1--23}
    (\bibinfo{year}{2012}).

  \bibitem{Filimonov_Sornette:2012}
    \bibinfo{author}{Filimonov, D.} \& \bibinfo{author}{Sornette, D.}
    \newblock \bibinfo{title}{Quantifying reflexivity in financial markets: Toward
    a prediction of flash crashes}.
    \newblock \emph{\bibinfo{journal}{Physical Review E}}
    \textbf{\bibinfo{volume}{85}}, \bibinfo{pages}{056108--1--9}
    (\bibinfo{year}{2012}).

  \bibitem{Bacry_etal:2013}
    \bibinfo{author}{Bacry, E.}, \bibinfo{author}{Delattre, S.},
    \bibinfo{author}{Hoffmann, M.} \& \bibinfo{author}{Muzy, J.~F.}
    \newblock \bibinfo{title}{Modelling microstructure noise with mutually exciting
    point processes}.
    \newblock \emph{\bibinfo{journal}{Quantitative Finance}}
    \textbf{\bibinfo{volume}{13}}, \bibinfo{pages}{65--77}
    (\bibinfo{year}{2013}).

  \bibitem{hardiman2013critical}
    \bibinfo{author}{Hardiman, S.~J.}, \bibinfo{author}{Bercot, N.} \&
    \bibinfo{author}{Bouchaud, J.-P.}
    \newblock \bibinfo{title}{Critical reflexivity in financial markets: a hawkes
    process analysis}.
    \newblock \emph{\bibinfo{journal}{The European Physical Journal B}}
    \textbf{\bibinfo{volume}{86}}, \bibinfo{pages}{1--9} (\bibinfo{year}{2013}).

  \bibitem{rambaldi2014modeling}
    \bibinfo{author}{Rambaldi, M.}, \bibinfo{author}{Pennesi, P.} \&
    \bibinfo{author}{Lillo, F.}
    \newblock \bibinfo{title}{Modeling fx market activity around macroeconomic
    news: a hawkes process approach}.
    \newblock \emph{\bibinfo{journal}{Available at ArXiV 1405.6047}}
    (\bibinfo{year}{2014}).

  \bibitem{Bacry_etal:2015}
    \bibinfo{author}{Bacry, E.}, \bibinfo{author}{Mastromatteo, I.} \&
    \bibinfo{author}{Muzy, J.~F.}
    \newblock \bibinfo{title}{{H}awkes processes in finance}.
    \newblock \emph{\bibinfo{journal}{Available at ArXiV 1502.04592}}
    (\bibinfo{year}{2015}).

  \bibitem{Barndorff-Nielsen_Shephard:2004}
    \bibinfo{author}{Barndorff-Nielsen, O.~E.} \& \bibinfo{author}{Shephard, N.}
    \newblock \bibinfo{title}{Power and bipower variation with stochastic
    volatility and jumps}.
    \newblock \emph{\bibinfo{journal}{Journal of Financial Econometrics}}
    \textbf{\bibinfo{volume}{2}}, \bibinfo{pages}{1--48} (\bibinfo{year}{2004}).

  \bibitem{Corsi_etal:2010}
    \bibinfo{author}{Corsi, F.}, \bibinfo{author}{Pirino, D.} \&
    \bibinfo{author}{Ren\`o, R.}
    \newblock \bibinfo{title}{Threshold bipower variation and the impact of jumps
    on volatility forecasting}.
    \newblock \emph{\bibinfo{journal}{Journal of Econometrics}}
    \textbf{\bibinfo{volume}{159}}, \bibinfo{pages}{276--288}
    (\bibinfo{year}{2010}).

  \bibitem{Joulin_etal:2008}
    \bibinfo{author}{Joulin, A.}, \bibinfo{author}{Lefevre, A.},
    \bibinfo{author}{Grunberg, D.} \& \bibinfo{author}{Bouchaud, J.-P.}
    \newblock \bibinfo{title}{Stock price jumps: {N}ews and volume play a minor
  role}.
  \newblock \emph{\bibinfo{journal}{Wilmott Magazine}}
  \textbf{\bibinfo{volume}{Sep/Oct}}, \bibinfo{pages}{1--7}
  (\bibinfo{year}{2008}).

\bibitem{Vere-Jones:1970}
  \bibinfo{author}{Vere-Jones, D.}
  \newblock \bibinfo{title}{Stochastic models for earthquake occurrence}.
  \newblock \emph{\bibinfo{journal}{Journal of the Royal Statistical Society,
  Series B}} \textbf{\bibinfo{volume}{32}}, \bibinfo{pages}{1--62}
  (\bibinfo{year}{1970}).

\bibitem{Vere-Jones_Ozaki:1982}
  \bibinfo{author}{Vere-Jones, D.} \& \bibinfo{author}{Ozaki, T.}
  \newblock \bibinfo{title}{Some examples of statistical inference applied to
  earthquake data}.
  \newblock \emph{\bibinfo{journal}{Annals of the Institute of Statistical
  Mathematics}} \textbf{\bibinfo{volume}{34}}, \bibinfo{pages}{189--207}
  (\bibinfo{year}{1982}).

\bibitem{Ogata:1988}
  \bibinfo{author}{Ogata, Y.}
  \newblock \bibinfo{title}{Statistical models for earthquake occurrences and
  residual analysis for point processes}.
  \newblock \emph{\bibinfo{journal}{Journal of the American Statistical
  Association}} \textbf{\bibinfo{volume}{83}}, \bibinfo{pages}{9--27}
  (\bibinfo{year}{1988}).

\bibitem{Daley_Vere-Jones:2003}
  \bibinfo{author}{Daley, D.~J.} \& \bibinfo{author}{Vere-Jones, D.}
  \newblock \emph{\bibinfo{title}{An Introduction to the Theory of Point
  Processes Volume I: Elementary Theory and Methods}}
  (\bibinfo{publisher}{Springer, Heidelberg}, \bibinfo{year}{2003}).

\bibitem{Bauwens_Hautsch:2009}
  \bibinfo{author}{Bauwens, L.} \& \bibinfo{author}{Hautsch, N.}
  \newblock \bibinfo{title}{Modelling financial high frequency data using point
  processes}.
  \newblock In \bibinfo{editor}{Mikosch, T.}, \bibinfo{editor}{Kreiss, J.-P.},
  \bibinfo{editor}{Davis, R.~A.} \& \bibinfo{editor}{Andersen, T.~G.} (eds.)
  \emph{\bibinfo{booktitle}{Handbook of Financial Time Series}},
  \bibinfo{pages}{953--979} (\bibinfo{publisher}{Springer, Berlin},
  \bibinfo{year}{2009}).

\end{thebibliography}
\end{document}